\documentclass
[aps,prl,twocolumn,showpacs,lengthcheck,preprintnumbers,superscriptaddress]{revtex4}%
\usepackage{amsfonts}
\usepackage{amsmath}
\usepackage{amssymb}
\usepackage{graphicx}%
\providecommand{\U}[1]{\protect\rule{.1in}{.1in}}
\begin{document}
\preprint{J. Phys. A: Math. Theor. \textbf{47}, 245202 (2014)}
\title{Accurate one-dimensional effective description of realistic matter-wave gap solitons}
\author{A. Mu\~{n}oz Mateo}
\affiliation{Departamento de F\'{\i}sica, Facultad de F\'{\i}sica, Universidad de La
Laguna, E-38206 La Laguna, Tenerife, Spain}
\affiliation{Centre for Theoretical Chemistry and Physics and New Zealand Institute for
Advanced Study, Massey University, Private Bag 102904 NSMC, Auckland 0745, New Zealand}

%\altaffiliation[Present address: ]{New Zealand Institute for Advanced Study, Massey University, Auckland, New Zealand}

\author{V. Delgado}
\affiliation{Departamento de F\'{\i}sica, Facultad de F\'{\i}sica, Universidad de La
Laguna, E-38206 La Laguna, Tenerife, Spain}


\begin{abstract}
We consider stationary matter-wave gap solitons realized in Bose--Einstein
condensates loaded in one-dimensional (1D) optical lattices and
investigate whether the effective 1D equation proposed in [Phys. Rev. A
\textbf{77}, 013617 (2008)] can be a reliable alternative to the
three-dimensional treatment of this kind of system in terms of the
Gross--Pitaevskii equation (GPE). Our results demonstrate that, unlike the
standard 1D GPE (which is not applicable in most realistic situations), the
above effective model is able to correctly predict the distinctive
trajectories characterizing the different gap soliton families as well as the
corresponding axial wavefunctions along the entire band gaps. It can also
predict the stability properties of the different gap soliton families as
follows from both a linear stability analysis and a representative set of
numerical computations. In particular, by numerically solving the 
corresponding Bogoliubov--de Gennes equations we show that the effective 1D 
model gives the correct spectrum of complex eigenfrequencies responsible for 
the dynamical stability of the system, thus providing us with a useful tool 
for the physical description of stationary matter-wave gap solitons in 1D 
optical lattices.

\end{abstract}
\maketitle

%\email{ammateo@ull.es}

%\email{vdelgado@ull.es}

\section{I. INTRODUCTION}

Bose--Einstein condensates (BECs) of dilute atomic gases are an invaluable
source of insight into the quantum behavior of matter at macroscopic scales
\cite{PeSmB,PiStB}. Topics like vortex structures or nonlinear matter waves
are being elucidated through BEC investigations \cite{Carre1}. In particular,
considerable effort has been invested in recent years in the study of solitary
matter waves (solitons) \cite{RevDS,Kon1,Mor1} due, in part, to their
similarity with well-known equivalent structures in the field of nonlinear
optics \cite{KivGOP,Kiv2D,RMP-Malomed}. A great variety of solitons have been
experimentally observed on the background of condensed atomic clouds. Dark
solitons have been found \cite{Burger1,Phil1,Hau1,Corn1,DS-2} as excitations 
in systems with repulsive interatomic interactions (the analogue of
self-defocusing nonlinear media), while bright solitons have been generated
\cite{Hul1,Salo1,Wie1} as ground states when the interaction is attractive
(self-focusing nonlinear media). A particularly important special case of
bright solitons is that of gap solitons \cite{Mor1}. These matter-wave
nonlinear structures are localized states that can occur in BECs with
repulsive interatomic interactions if a periodic potential (such as that
produced by an optical lattice) is present. Although gap solitons have been
observed and studied in nonlinear optics since some time ago, only more
recently have they been found in BEC experiments \cite{Ober1}. This has
triggered a renewed theoretical interest in the study of these nonlinear 
structures.

Strictly speaking, solitons in a one-dimensional (1D) setting are exact 
solutions of the one-dimensional nonlinear Schr\"{o}dinger equation 
(1D NLSE).
Accordingly, in the framework of the mean-field approximation, matter-wave
solitonic structures in 1D optical lattices have usually been investigated
theoretically by means of the 1D Gross--Pitaevskii equation (GPE)
\cite{GS1D,Kiv1,Efre1,Abd1,Mal1,Wu1}, which is nothing but a nonlinear
Schr\"{o}dinger equation describing the evolution of the condensate
wavefunction in the presence of the confining potentials. However,
soliton-like states realized in BECs always have an intrinsic
three-dimensional (3D) character, even when the trapping potential imposes 
a tight radial confinement. In fact, typical experimental parameters for the
generation of gap solitons in BECs lie in a region of parameter space where
the quasi-1D approximation is not strictly valid \cite{Mor1,Ober1} and, as a
consequence, very often the 1D GPE cannot quantitatively reproduce the 
dynamics of realistic matter-wave gap solitons.

%============================================================================

Various approaches have been put forward with the aim of incorporating in the
description of condensed gases in elongated geometries the consequences of
their intrinsic 3D character. Effective models for studying the stationary 
properties of highly anisotropic BECs beyond the analytically solvable
perturbative and Thomas--Fermi regimes were proposed in Refs. \cite{Das1,VDB1,VDB2}.
An analytic expression for the local chemical potential $\mu_{\bot}\!(n_{1}\!)$ of 
elongated condensates as a function of the axial linear density was independently 
proposed in Refs. \cite{VDB2} and \cite{Fuch1}. 
Such expression remains valid in the 3D--1D crossover regime and, in the case of 
\cite{VDB2}, is also valid in the presence of an axially symmetric vortex. 
Effective equations have also been introduced to study topics such as solitons 
\cite{Jack1}, the influence of the transverse confining on the dynamics of BECs 
in 1D optical lattices \cite{Yuka1}, or the emergence of Faraday waves 
\cite{Nico3}. Of particular relevance because of their simplicity and usefulness are 
the effective 1D equations derived in Refs. \cite{Salas1} and \cite{Ef1DEqs,VDB2}, 
which govern the condensate dynamics accounting properly for the contributions from 
excited transversal modes through a nonpolynomial nonlinear term. Several variants
of the above equations aimed at improving their accuracy to the price of 
incorporating adjustable \cite{Adhik1} or a larger number of variational parameters 
\cite{Kam1,Clark1} have also been proposed. However, the intended increase in 
accuracy hardly compensates for the complexity of the resulting equations or the 
arbitrariness in the choice of free parameters.
On the other hand, a number of generalizations of the above equations have been 
derived to study issues such as the expansion of elongated BECs \cite{Modug1}, 
spin-1 atomic condensates \cite{You1}, matter-wave vortices \cite{Salas3}, BECs in 
anisotropic transverse traps \cite{Salas4}, binary condensates in mixed dimensions 
\cite{Adhik2}, dark--bright solitons in two-component BECs \cite{Carre3}, dipolar 
BECs \cite{Adhik4}, or localized modes in condensates with spin-orbit and Rabi 
couplings \cite{Salas6}.
The equations of Refs. \cite{Salas1} and \cite{Ef1DEqs} have also been used to 
study BECs in nonlinear lattices \cite{Salas5}, bright solitons in condensates with 
inhomogeneous defocusing nonlinearities \cite{Mal2}, or the Bogoliubov spectrum of 
elongated condensates \cite{Ben1}.
Since these effective equations are extensions of the standard
1D GPE (to which they reduce in the proper limit) and can be numerically
solved with similar computational effort, they may offer a clear advantage
over the latter. In this work, we will focus on the effective equation
proposed in Refs. \cite{Ef1DEqs,VDB2}. This equation has been used in the
description of matter-wave dark solitons in elongated BECs and has proven to
give results in good agreement with the experiments 
\cite{Kev1,Wel1,Kev2,Isa1}. As was demonstrated in Ref. \cite{VDB7}, it can 
also be applied to the study of stationary gap solitons in the different 
physically relevant mean-field regimes. 
This is a nontrivial issue because this kind of systems are characterized by a
small spatial scale (the lattice period $d$) that can represent a rather
restrictive condition. However, to establish the utility of the model in
realistic situations and determine whether it may be a real alternative to a
fully 3D treatment one has to analyze its ability to reproduce a number of
physical properties of fundamental interest in the theoretical description of
stationary matter-wave gap solitons that were not studied in Ref. \cite{VDB7}.
One of these properties is the dependence of the soliton chemical potential
$\mu$ on the number $N$ of constituent atoms, a quantity that can be used to
define trajectories in the $\mu-N$ plane which are essential for the
classification of gap solitons into different families. Another one is the
frequency spectrum of the elementary excitations determining the dynamical
stability of the gap soliton family. To provide an answer to the previous
question, in this work we analyze to what extent the 1D GPE and the effective
equation of Refs. \cite{Ef1DEqs,VDB2} can reproduce the above-mentioned 
fundamental physical properties in realistic situations. To this end we 
consider condensed atomic clouds confined by transverse harmonic potentials 
in a range of parameters typical of current experiments and compare with 
exact numerical results obtained from the full 3D GPE.

Since in this work we are interested in the mean-field regime, we shall 
restrict our analysis to physical configurations with a large number of atoms 
per lattice well, which are amenable to a treatment in terms of a mean-field 
macroscopic wavefunction satisfying the GPE.

%============================================================================

The paper is organized as follows. Section II briefly reviews the theoretical
models considered in this work. In Secs. III and IV we analyze the numerical
results: in Sec. III we focus on the wavefunctions and the distinctive
trajectories of the different gap soliton families, and in Sec. IV we study
the frequency spectrum of the elementary excitations and the stability 
properties of these nonlinear structures. Finally, in Sec. V we summarize our 
results and present our conclusions.

\section{II. EFFECTIVE MODELS FOR GAP SOLITONS IN 1D OPTICAL LATTICES}

When a periodic potential is imposed on a BEC, the chemical potential of the
system develops a band-gap structure. In these circumstances, it is possible
to find either extended states with chemical potential inside the bands
(nonlinear Bloch waves) or localized gap solitons between bands. At zero
temperature and in the mean-field regime, these coherent states are accurately
described by the 3D GPE \cite{PeSmB,PiStB}:
\begin{equation}
i\hbar\frac{\partial\psi}{\partial t}=\left(  -\frac{\hbar^{2}}{2m}\nabla
^{2}+V(\mathbf{r})+gN\left\vert \psi\right\vert ^{2}\right)  \psi,
\label{3DGPE}%
\end{equation}
where $g=4\pi\hbar^{2}a/m$ is the interaction parameter, with $a$ being the
\textsl{s}-wave scattering length, $N$ is the number of atoms, and
$V(\mathbf{r})$ is the potential of the trap. In this work we will consider an
external potential of the form $V(\mathbf{r})=V_{\bot}(r_{\bot})+V_{z}(z)$,
which besides the axial term $V_{z}(z)=V_{0}\sin^{2}(\pi z/d)$ accounting for
a 1D optical lattice with period $d$, includes a transverse harmonic trapping
$V_{\bot}(r_{\bot})=m\omega_{\bot}^{2}r_{\bot}^{2}/2$, where $r_{\bot}%
^{2}=x^{2}+y^{2}$. 

A description in terms of the GPE corresponds to a classical mean-field 
description that neglects quantum correlations. It represents the classical 
limit of the underlying quantum field theory and becomes an excellent 
approximation in the limit $N \rightarrow \infty$ with the nonlinear 
interaction parameter $gN$ held constant.
In practice, quantum fluctuations decay rapidly with $N$ and one expects the 
GPE to be a good approximation for configurations with a few hundreds of atoms 
per lattice well.

Although the 3D GPE can be solved numerically with no
approximations \cite{PRL06,Mott}, it demands a considerable computational
effort when either the 3D character of the system or the nonlinearity
increases. In order to simplify the computational problem, simpler models with
reduced dimensionality are commonly used in the study of gap solitons in 1D
optical lattices. Whenever it is possible to make the assumption that the
condensate remains in the ground state of the transverse harmonic trap (so
that higher transverse modes are not excited), a good approximation to Eq.
(\ref{3DGPE}) is given by the 1D GPE with rescaled nonlinearity:
\begin{equation}
i\hbar\frac{\partial\phi}{\partial t}=-\frac{\hbar^{2}}{2m}\frac{\partial
^{2}\phi}{\partial z^{2}}+V_{z}(z)\phi+g_{1D}N\left\vert \phi\right\vert
^{2}\phi, \label{1DGPE}%
\end{equation}
where $g_{1D}=2a\hbar\omega_{\bot}$ is the 1D interaction parameter, obtained
after averaging over the `frozen' transverse degrees of freedom, and 
$\phi(z)$ is the axial wavefunction of
the condensate. It is important to bear in mind that this equation is only
valid when the axial energies are small enough in comparison with the radial
quantum, $\hbar\omega_{\bot}$, a condition that, in general, is not satisfied
in realistic situations.

A simple dimensional analysis shows that the physical problem is characterized
by three dimensionless parameters which can be conveniently chosen as%
\begin{equation}
\frac{E_{R}}{\hbar\omega_{\bot}},\;\frac{V_{0}}{E_{R}},\;\frac{Na}%
{d} \label{con1}%
\end{equation}
where $E_{R}\equiv\hbar^{2}k^{2}/2m$ is the recoil energy of the lattice, with
$k=\pi/d$ being the lattice wavevector, $s=V_{0}/E_{R}$ is the lattice depth,
and $Na/d$ is the nonlinear coupling constant determining the mean-field
interaction energy. The derivation of Eq. (\ref{1DGPE}) relies on the
assumption that the 3D wavefunction $\psi$ can be factorized in independent
radial and axial components, with the radial component corresponding to the
Gaussian wavefunction of the ground state. Such assumption entails an implicit
adiabatic approximation according to which the transverse degrees of freedom
can instantaneously follow the axial dynamics. In order for this to be true,
the characteristic time scale of the axial motion must be much longer than
that of the radial motion ($\sim\omega_{\bot}^{-1}$). Moreover, since the
validity of the 1D GPE (\ref{1DGPE}) requires the radial configuration to
remain in the ground state, the axial energies involved must be sufficiently
small in comparison with the characteristic energy of the transverse harmonic
oscillator, $\hbar\omega_{\bot}$. Taking into account that the axial energy of
the underlying linear problem is of the order of the recoil energy $E_{R}$ and
that the nonlinear interaction energy is of the order of $an_{1}\hbar
\omega_{\bot}$ (with $n_{1}=N/d$ being the linear density of the gap soliton
along the axial direction), the above requirements can be stated as%
\begin{equation}
E_{R}\ll\hbar\omega_{\bot},\;\;an_{1}\ll1.\label{con2}%
\end{equation}
These conditions, which ensure the permanence of the system in the transverse
ground state, also guarantee automatically the validity of the adiabatic
approximation. However, in order for the mean-field approximation implicit in
Eqs. (\ref{3DGPE}) and (\ref{1DGPE}) to be applicable, it is also necessary
that $N\gg1$. Incorporating this latter requirement, the conditions for the
validity of the 1D GPE can be rewritten as%
\begin{equation}
d\gg a_{\bot},\;\;1\ll N\ll d/a,\label{con3}%
\end{equation}
where $a_{\bot}=\sqrt{\hbar/m\omega_{\bot}}$ is the radial harmonic-oscillator
length. These conditions are very restrictive and usually are not met in real
condensates. For instance, for $^{87}$Rb atoms $a=5.3$ $%
%TCIMACRO{\unit{nm}}%
%BeginExpansion
\operatorname{nm}%
%EndExpansion
$ and $a_{\bot}=10.78/\sqrt{\omega_{\bot}/2\pi}$ $%
%TCIMACRO{\unit{\U{3bc}m}}%
%BeginExpansion
\operatorname{\mu m}%
%EndExpansion
$ (with $\omega_{\bot}/2\pi$ given in $%
%TCIMACRO{\unit{Hz}}%
%BeginExpansion
\operatorname{Hz}%
%EndExpansion
$) while, typically, the lattice parameter $d$ varies between $0.4$ and $1.6$
$%
%TCIMACRO{\unit{\U{3bc}m}}%
%BeginExpansion
\operatorname{\mu m}%
%EndExpansion
$ \cite{Arim1}, the lattice depth $s$ between $0$ and $30$ \cite{Ingus1} and,
for elongated condensates, $\omega_{\bot}/2\pi$ varies between $85$ and $1000$
$%
%TCIMACRO{\unit{Hz}}%
%BeginExpansion
\operatorname{Hz}%
%EndExpansion
$. Thus, typically, $d\lesssim1.6$ $%
%TCIMACRO{\unit{\U{3bc}m}}%
%BeginExpansion
\operatorname{\mu m}%
%EndExpansion
$, $a_{\bot}\gtrsim0.34$ $%
%TCIMACRO{\unit{\U{3bc}m}}%
%BeginExpansion
\operatorname{\mu m}%
%EndExpansion
$, and $d/a\lesssim300$ which makes the conditions (\ref{con3}) hard to
satisfy. In particular, in the experiment of Ref. \cite{Ober1}, where gap
solitons were observed in a $^{87}$Rb condensate, $d=0.39$ $%
%TCIMACRO{\unit{\U{3bc}m}}%
%BeginExpansion
\operatorname{\mu m}%
%EndExpansion
$, $a_{\bot}=1.17$ $%
%TCIMACRO{\unit{\U{3bc}m}}%
%BeginExpansion
\operatorname{\mu m}%
%EndExpansion
$, $s=0.7$, and $N\sim900$ atoms. In that of Ref. \cite{Ingus1}, $d=0.42$ $%
%TCIMACRO{\unit{\U{3bc}m}}%
%BeginExpansion
\operatorname{\mu m}%
%EndExpansion
$, $a_{\bot}=1.14$ $%
%TCIMACRO{\unit{\U{3bc}m}}%
%BeginExpansion
\operatorname{\mu m}%
%EndExpansion
$, $s=0-30$, and $N\sim3\times10^{5}$ $^{87}$Rb\ atoms. It is important to
note, however, that even though most experiments have been carried out in the
above parameter regimes, larger values for $d$ and $\omega_{\bot}$ (which can
realize the 1D mean-field regime) are well within experimental reach.

Next, we briefly recall the 1D effective model proposed in 
Refs. \cite{Ef1DEqs,VDB2}, which is an extension of the standard 1D GPE. 
As before, the starting point is the adiabatic approximation, which allows the 
factorization of the 3D wavefunction in the form%
\begin{equation}
\psi(\mathbf{r},t)=\varphi(\mathbf{r}_{\bot};n_{1})\phi(z,t), \label{II-4}%
\end{equation}
where the radial wavefunction $\varphi$ is assumed to be normalized to unity
and $n_{1}$ is the local linear density%
\begin{equation}
n_{1}(z,t)=N|\phi(z,t)|^{2}=N\int d^{2}\mathbf{r}_{\bot}|\psi(\mathbf{r}%
_{\bot},z,t)|^{2}. \label{II-5}%
\end{equation}
After substituting Eq. (\ref{II-4}) in Eq. (\ref{3DGPE}) and averaging over
the radial degrees of freedom one obtains the following 1D effective equation%
\begin{equation}
i\hbar\frac{\partial\phi}{\partial t}\!=\!-\frac{\hbar^{2}}{2m}\frac
{\partial^{2}\phi}{\partial z^{2}}+V_{z}\phi+\mu_{\bot}\phi, \label{II-6}%
\end{equation}
where $\mu_{\bot}\!(n_{1}\!)$ is the transverse local chemical potential which
follows from the stationary 2D GPE%
\begin{equation}
\left(  \!-\frac{\hbar^{2}}{2m}\nabla_{\bot}^{2}+V_{\bot}(\mathbf{r}_{\bot
})+gn_{1}\left\vert \varphi\right\vert ^{2}\!\right)  \varphi=\mu_{\bot}%
(n_{1})\varphi. \label{II-7}%
\end{equation}
When the dimensionless linear density is small enough ($an_{1}\ll1$), this
latter equation can be solved perturbatively and, to the lowest order
(\textit{i.e.}, for $\varphi(\mathbf{r}_{\bot})$ given by the Gaussian ground
state of the corresponding harmonic oscillator), one obtains $\mu_{\bot}%
=\hbar\omega_{\bot}(1+2an_{1})$. Substituting back in Eq. (\ref{II-6}) and
taking into account Eq. (\ref{II-5}) it is thus clear that one arrives at the
1D GPE (\ref{1DGPE}). However, a more general solution to Eq. (\ref{II-7}) can
be found. Using two independent approaches it was demonstrated in Refs.
\cite{VDB2,Ef1DEqs} that, for condensates with no vorticity (the case we are
interested in here), a very accurate solution, valid for arbitrary $an_{1}$,
is given by%
\begin{equation}
\mu_{\bot}=\hbar\omega_{\bot}\sqrt{1+4an_{1}}. \label{II-8}%
\end{equation}
Substituting this expression in Eq. (\ref{II-6}), one finally finds%
\begin{equation}
i\hbar\frac{\partial\phi}{\partial t}=-\frac{\hbar^{2}}{2m}\frac{\partial
^{2}\phi}{\partial z^{2}}+V_{z}(z)\phi+\hbar\omega_{\bot}\sqrt{1+4aN\left\vert
\phi\right\vert ^{2}}\phi. \label{E1DE}%
\end{equation}
This effective 1D equation governs the axial dynamics of the condensate,
having a much wider range of applicability than the standard 1D GPE
(\ref{1DGPE}). In particular, now the linear density $an_{1}$ can take
arbitrary values and in the limit $an_{1}\ll1$ one recovers Eq. (\ref{1DGPE}).
Yet, the applicability of the above equation still requires the fulfillment of
the adiabatic approximation and, as far as the dynamical problem is concerned,
this condition can be hard to satisfy in the presence of an optical lattice.
Things are different, however, for the stationary problem, which is, in fact,
the most relevant in the characterization of matter-wave gap solitons. In this
case, the time scale of the axial dynamics tends to infinity and the adiabatic
approximation always holds. As a result, unlike the usual 1D GPE
(\ref{1DGPE}), the effective model (\ref{E1DE}) will be able to correctly
predict the trajectories in $\mu-N$ plane as well as the wavefunctions and
stability properties of the different gap soliton families.

%%%%%%%%%%%%%%%%%%%%%%%%%%%%%%%%%%%%%%%%%%%%%%%%%%%%%%%%%%%%%%%%%%%%%%%%%%%%%%

\begin{figure}[t]
\begin{center}
\includegraphics[width=7.3cm]{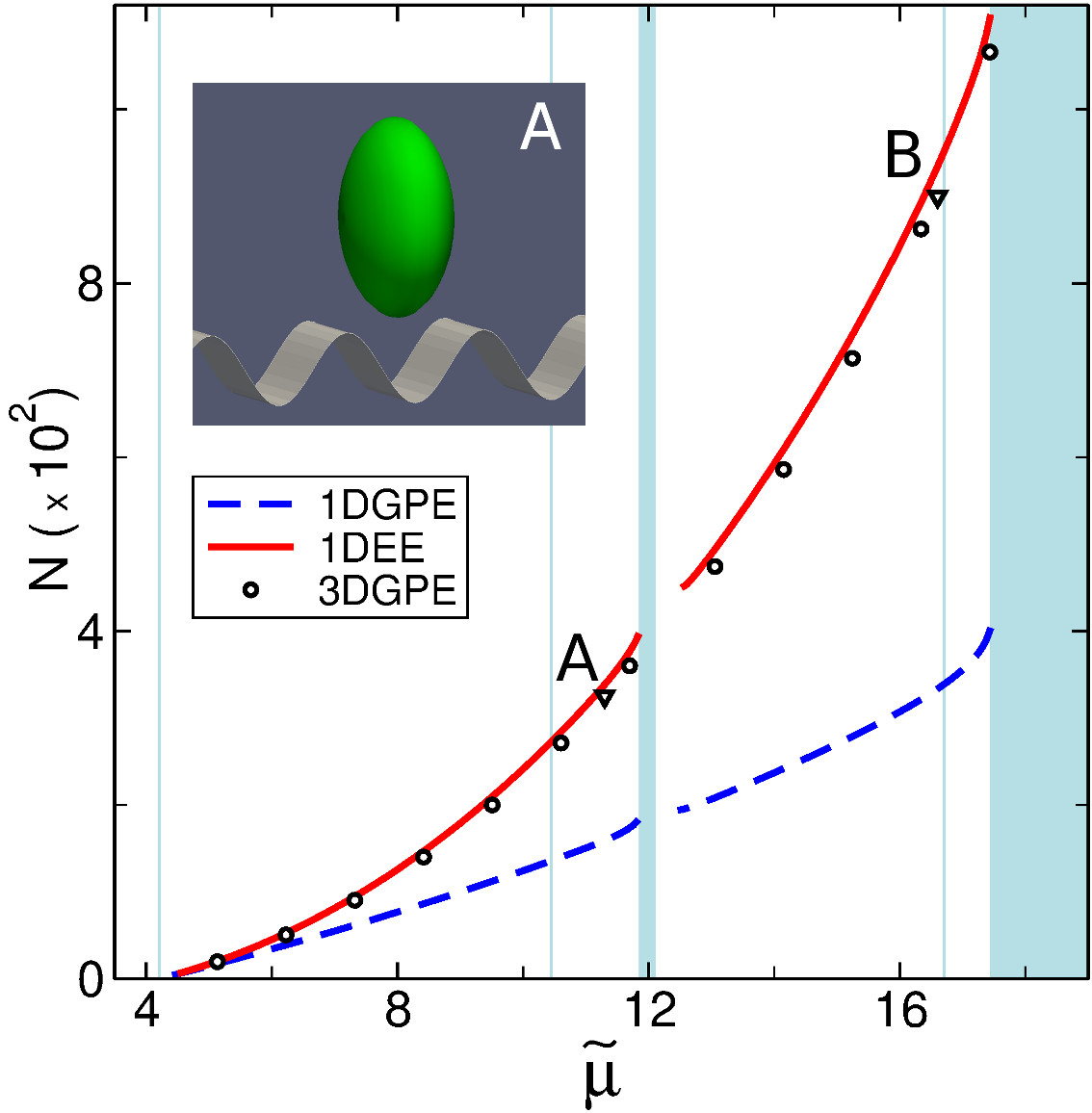}
\end{center}
\caption{(Color online) The family of fundamental gap solitons as predicted by
the different 1D effective models. $N$ represents the number of $^{87}$Rb
atoms, and $\tilde{\mu}\equiv(\mu-\hbar\omega_{\bot})/E_{R}$ is the
dimensionless chemical potential. The solid red curve is the prediction from
the 1D effective equation (\ref{E1DE}), the dashed blue curve is that from the
1D GPE (\ref{1DGPE}), and the open symbols are exact results obtained by
solving numerically the full 3D GPE (\ref{3DGPE}). The inset shows a
representative example of a gap soliton in this family (point A) in terms of a
density isosurface (corresponding to $5\%$ of its maximum value). The local
phase (which is uniform in this case) is also represented as a color map.}%
\label{Fig1}%
\end{figure}

%%%%%%%%%%%%%%%%%%%%%%%%%%%%%%%%%%%%%%%%%%%%%%%%%%%%%%%%%%%%%%%%%%%%%%%%%%%%%%

\section{III. GAP SOLITON FAMILIES}

Next, we will apply both Eqs. (\ref{1DGPE}) and (\ref{E1DE}) to the study of
stationary gap solitons in a regime corresponding to typical experimental
parameters. To this end, by using a Newton continuation method we look for
numerical solutions of the form $\phi(z,t)=\phi_{0}(z)e^{-i\mu t/\hbar}$, with
$\mu$ being the condensate chemical potential. These solutions define
distinctive trajectories in $\mu-N$ plane which, as already said, are
essential for the classification of gap solitons into different families. To
investigate how accurately the above 1D equations can predict such
trajectories and, thus, the physical properties of the various families, we
compare the predictions from both equations with those obtained from the
numerical solution of the full 3D GPE.

In what follows we consider a $^{87}$Rb condensate subject to a rather tight
transverse potential of frequency $\omega_{\bot}/2\pi=800$ $%
%TCIMACRO{\unit{Hz}}%
%BeginExpansion
\operatorname{Hz}%
%EndExpansion
$ and in the presence of a 1D optical lattice with period $d=1.5%
%TCIMACRO{\unit{\U{3bc}m}}%
%BeginExpansion
\operatorname{\mu m}%
%EndExpansion
$ and depth\ $s=$ $20$. For these parameters, the recoil energy of the lattice
takes the value $E_{R}=0.32\hbar\omega_{\bot}$.

%%%%%%%%%%%%%%%%%%%%%%%%%%%%%%%%%%%%%%%%%%%%%%%%%%%%%%%%%%%%%%%%%%%%%%%%%%%%%%

\begin{figure}[t]
\begin{center}
\includegraphics[width=8.5cm]{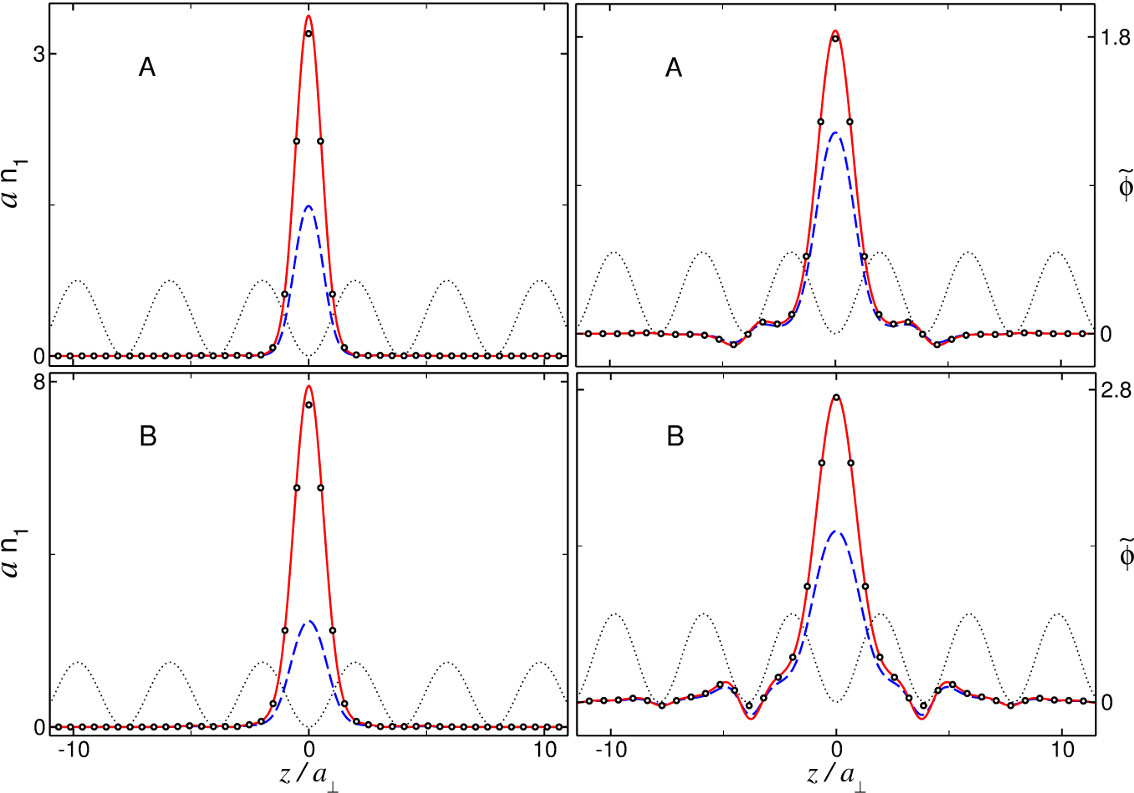}
\end{center}
\caption{(Color online) Axial densities (left panels) and wavefunctions (right
panels) of the fundamental gap solitons corresponding to points A and B in
Fig. \ref{Fig1}. Solid red lines are the predictions of the 1D effective
equation (\ref{E1DE}), dashed blue lines are those of the 1D GPE
(\ref{1DGPE}), and open symbols are exact results obtained by solving
numerically the full 3D GPE (\ref{3DGPE}).}%
\label{Fig2}%
\end{figure}

%%%%%%%%%%%%%%%%%%%%%%%%%%%%%%%%%%%%%%%%%%%%%%%%%%%%%%%%%%%%%%%%%%%%%%%%%%%%%%

Figure \ref{Fig1}\ shows the different theoretical predictions the above
equations make for the family of fundamental gap solitons. This figure depicts
the dimensionless chemical potential $\tilde{\mu}\equiv(\mu-\hbar\omega_{\bot
})/E_{R}$ of the solitons versus the number $N$ of constituent atoms. The
3D band-gap structure of the underlying linear problem is shown
in the background, with shaded stripes corresponding to the allowed energy
bands. The two narrow stripes located at $\tilde{\mu}=10.45$ and $16.7$ are
purely 3D Bloch bands associated with excited transversal states and, thus,
cannot be accounted for by any 1D model. They are replicas of the lowest
energy band (located at $\tilde{\mu}=4.2$) shifted up in energy, respectively,
by two and four quanta of the radial harmonic oscillator, $\hbar\omega_{\bot
}/E_{R}$ \cite{GS-3D,VDB8}. Open symbols in this figure are exact results
obtained by solving numerically the full 3D GPE (\ref{3DGPE}), the solid red
curve is the prediction from the 1D effective equation (\ref{E1DE}), and the
dashed blue curve is that from the standard 1D GPE (\ref{1DGPE}). As is
evident, this latter equation clearly fails. In fact, even in the first (1D)
gap, it makes errors larger than $100\%$ and thus can hardly be considered
valid. On the contrary, the effective model of Eq. (\ref{E1DE}) is in good
quantitative agreement with the\ 3D results over the entire curve. Gap
solitons in this family are nonlinear stationary states exhibiting a single
peak well localized at a lattice site. A representative example is displayed
in the inset, which shows the 3D atom density of a gap soliton of this family
containing $325$ particles (point A marked by an open triangle in the figure).
Its chemical potential, $\mu=4.7$ $\hbar\omega_{\bot}>\hbar\omega_{\bot}$,
reflects the fact that the corresponding transversal state is composed of
multiple harmonic-oscillator modes. The prediction the effective model
(\ref{E1DE}) makes for this soliton has an error in the number of particles of
only $2\%$. Figure \ref{Fig1} also shows that while the results from the 1D
GPE get worse as the number of atoms increase, those from the effective
equation (\ref{E1DE}) remain in good quantitative agreement with the 3D results.

%%%%%%%%%%%%%%%%%%%%%%%%%%%%%%%%%%%%%%%%%%%%%%%%%%%%%%%%%%%%%%%%%%%%%%%%%%%%%%

\begin{figure}[t]
\begin{center}
\includegraphics[width=7.5cm]{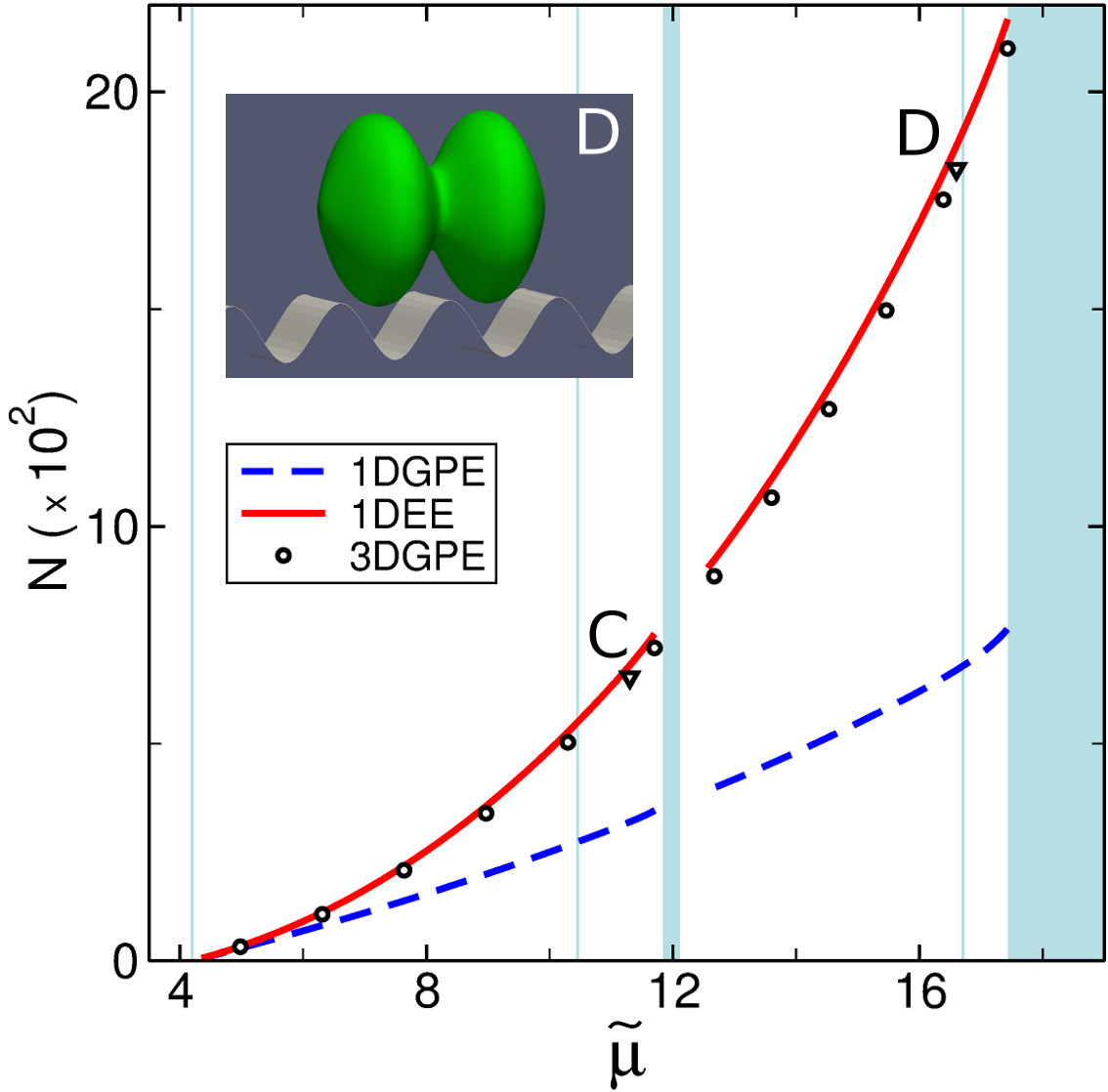}
\end{center}
\caption{(Color online) Same as Fig. \ref{Fig1}, but for the symmetric
composite family of gap solitons consisting of two in-phase peaks.}%
\label{Fig3}%
\end{figure}

%%%%%%%%%%%%%%%%%%%%%%%%%%%%%%%%%%%%%%%%%%%%%%%%%%%%%%%%%%%%%%%%%%%%%%%%%%%%%%

This behavior is corroborated by Fig. \ref{Fig2}. Panels A and B in this
figure display the dimensionless axial densities $an_{1}$ (left panels) and
wavefunctions $\tilde{\phi}(z)=\sqrt{aN}\phi_{0}(z)$ (right panels) of the
fundamental gap solitons corresponding, respectively, to points A and B in
Fig. \ref{Fig1}. As before, solid red curves have been obtained from the
effective equation (\ref{E1DE}), dashed blue curves from the standard 1D GPE,
and open symbols are numerical results obtained from the full 3D GPE. For
reference, the lattice potential is also shown in arbitrary units (dotted
lines). In order to deduce the axial wavefunction from the corresponding 3D
results we have used that $\tilde{\phi}(z)=\sqrt{an_{1}(z)}e^{i\varphi(z)}$,
where the linear density can be readily obtained by performing the integral in
Eq. (\ref{II-5}). As for the axial phase $\varphi(z)$, we have defined it as
the average local phase at every $z$-plane. Again the agreement between the
predictions of Eq. (\ref{E1DE}) and the 3D results is very good, while the 1D
GPE clearly fails.

\begin{figure}[t]
\begin{center}
\includegraphics[width=8.5cm]{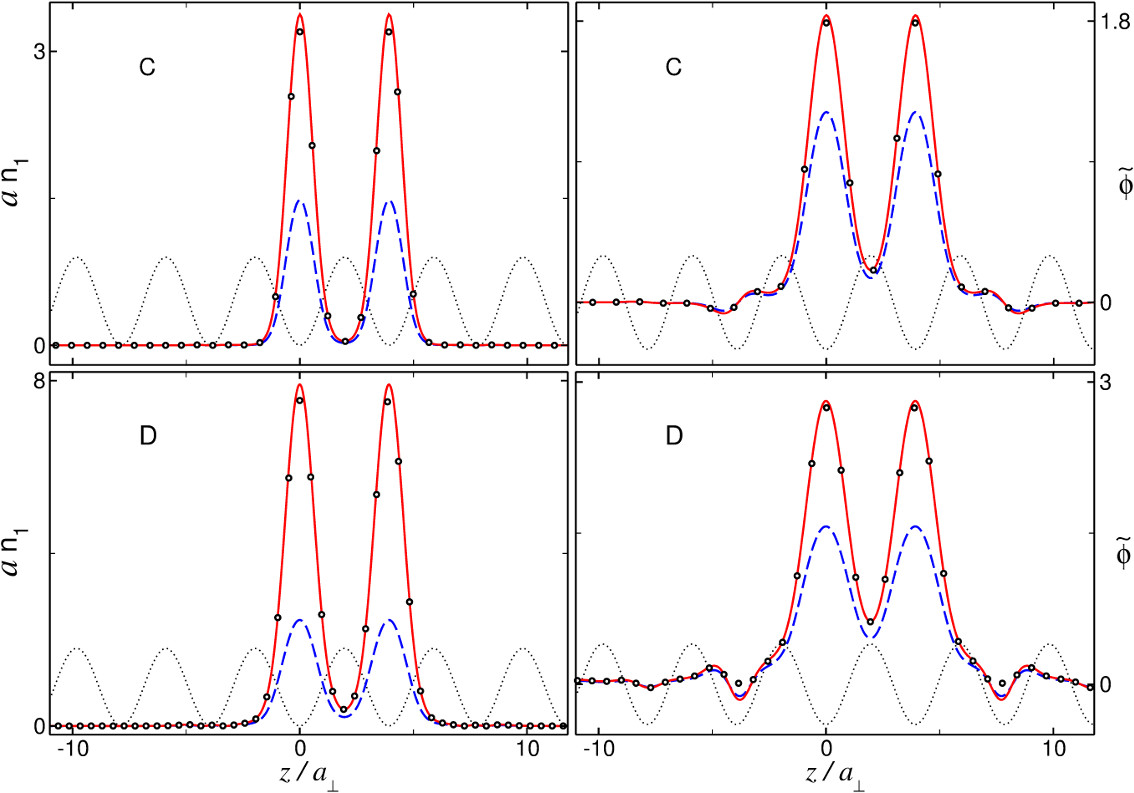}
\end{center}
\caption{(Color online) Same as Fig. \ref{Fig2}, but for the gap solitons
corresponding to points C and D in Fig. \ref{Fig3}.}%
\label{Fig4}%
\end{figure}

Similar conclusions can be drawn in the case of multiple gap solitons. Figure
\ref{Fig3} displays the trajectories in ${\mu}-N$ plane corresponding to the
composite family consisting of two in-phase peaks (see also Fig. \ref{Fig4}).
Solitons in this family are bound states formed by a symmetric superposition
of two fundamental gap solitons of the family $(1,0,0)$, \textit{i.e.} those
displayed in Fig. \ref{Fig1}. Here we are using the notation introduced in
Ref. \cite{GS-3D}, which enables us to classify the different fundamental gap
solitons into families characterized by the quantum numbers $(n,m,n_{r})$
corresponding to the 3D Bloch band from which they bifurcate, with
$n=1,2,3,\ldots$ being the band index of the corresponding 1D axial problem
and $m=0,\pm1,\pm2,\ldots$ and $n_{r}=0,1,2,\ldots$ being, respectively, the
angular-momentum and radial quantum numbers characterizing the transversal
state. In this work, as is usually the case, we restrict ourselves to gap
solitons consisting of fundamental constituents of the type $(n,0,0)$, which
are amenable to an effective treatment in terms of the 1D equations
(\ref{1DGPE}) and (\ref{E1DE}). Gap solitons with $(m,n_{r})\neq(0,0)$ have
been rarely considered in the literature \cite{GS-3D,VDB8}. They feature a
nontrivial radial topology and require a more elaborate treatment \cite{VDB8}.
The inset in Fig. \ref{Fig3} shows the atomic density (as an isosurface taken
at $5\%$ of the maximum density) of the soliton corresponding to point D
(marked by an open triangle). This soliton contains $1765$ atoms and has a
chemical potential $\tilde{\mu}=16.6$. As can be seen, the two fundamental
constituents occupy adjacent lattice sites and have the same phase (shown in
the inset in terms of a color map). This is also apparent from Fig.
\ref{Fig4}, which depicts the axial linear densities $an_{1}$ and
wavefunctions $\tilde{\phi}(z)$ of the gap solitons marked by C and D in Fig.
\ref{Fig3}. A simple comparison also shows that the wavefunctions of the
composite solitons are indeed symmetric superpositions of\ those in Fig.
\ref{Fig2}.

%%%%%%%%%%%%%%%%%%%%%%%%%%%%%%%%%%%%%%%%%%%%%%%%%%%%%%%%%%%%%%%%%%%%%%%%%%%%%%

%%%%%%%%%%%%%%%%%%%%%%%%%%%%%%%%%%%%%%%%%%%%%%%%%%%%%%%%%%%%%%%%%%%%%%%%%%%%%%

\begin{figure}[t]
\begin{center}
\includegraphics[width=7.5cm]{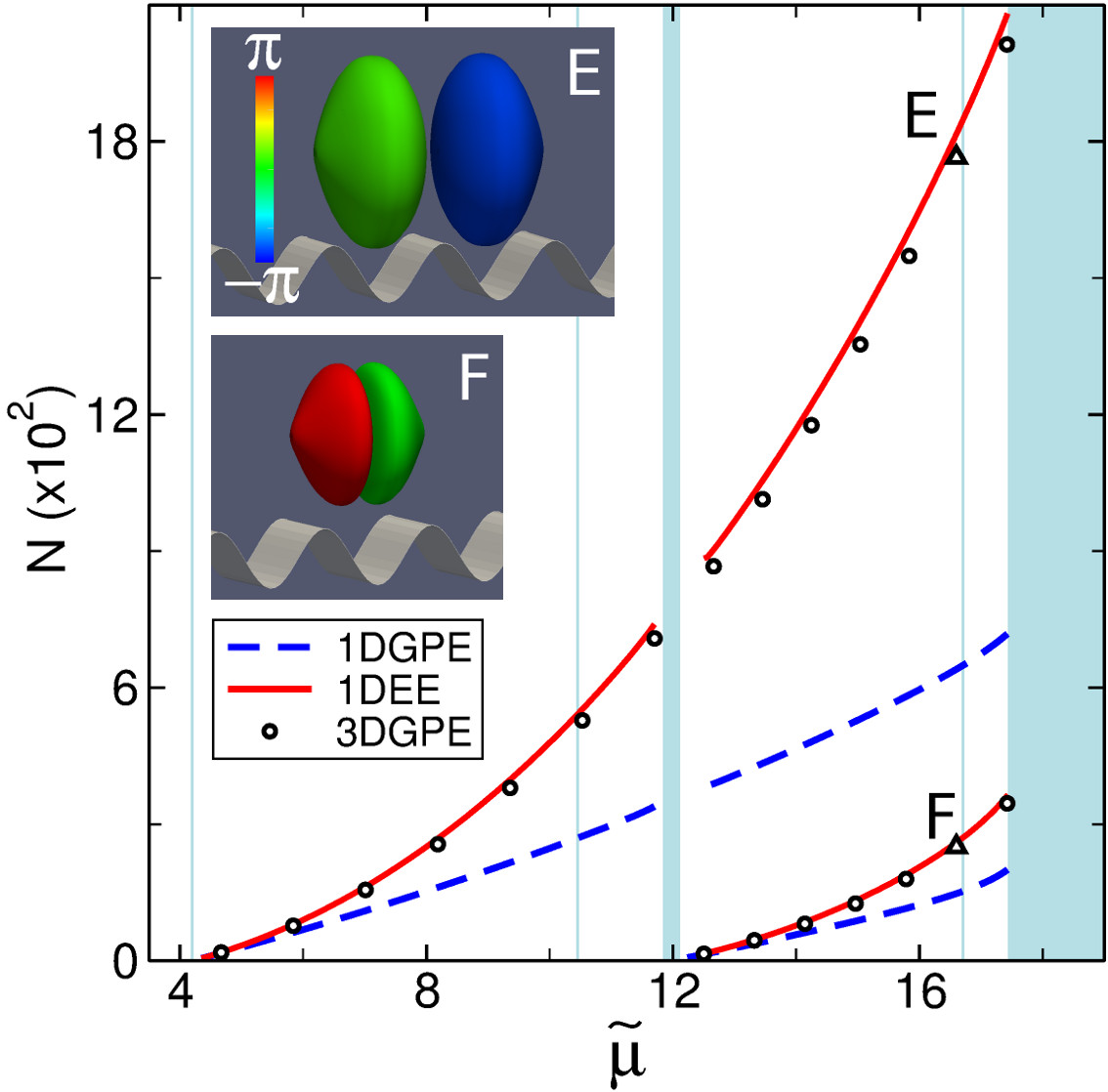}
\end{center}
\caption{(Color online) Same as Fig. \ref{Fig1}, but for the antisymmetric
composite (upper curves) and fundamental (lower curves) gap soliton families
consisting of two out-of-phase peaks.}%
\label{Fig5}%
\end{figure}

\begin{figure}[t]
\begin{center}
\includegraphics[width=8.5cm]{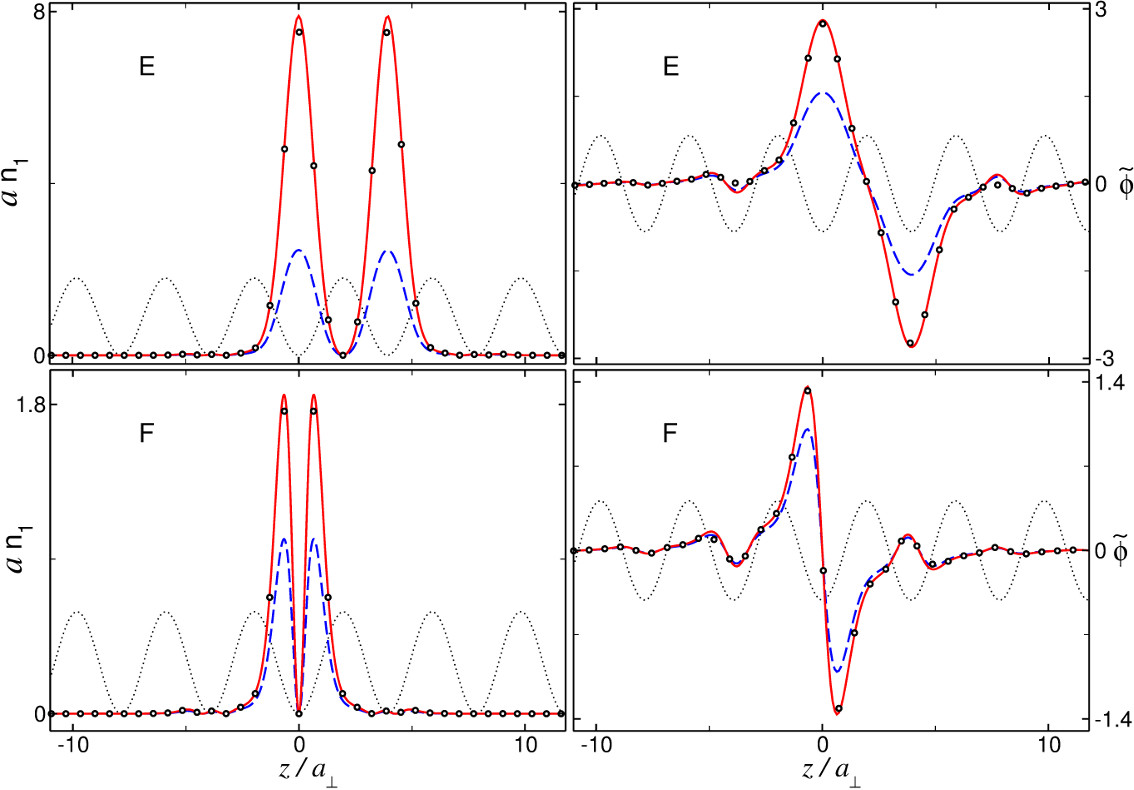}
\end{center}
\caption{(Color online) Same as Fig. \ref{Fig2}, but for the gap solitons
corresponding to points E and F in Fig. \ref{Fig5}.}%
\label{Fig6}%
\end{figure}

Gap solitons consisting of two out-of-phase peaks are analyzed in Fig.
\ref{Fig5}. The upper trajectories in this figure (those bifurcating from the
first Bloch band) correspond to composite solitons formed by an antisymmetric
superposition of two\ fundamental $(1,0,0)$ gap solitons. An example of a
soliton of this family is displayed in the upper inset, which shows a
representation of the 3D wavefunction of the soliton corresponding to point E
in the figure. This wavefunction is given in terms of an isosurface of the
atomic density (taken at $5\%$ of its maximum) and a color map indicating the
local phase at each point. As in the previous case, the two fundamental
constituents of this compound soliton occupy adjacent lattice sites, but now
they exhibit a $\pi$ jump in their relative phase. A detailed comparison of
the axial density and wavefunction of this soliton (given in the upper panels
of Fig. \ref{Fig6}) with those of the corresponding fundamental solitons
(given in the lower panels of Fig. \ref{Fig2}) confirms that this soliton is a
bound state consisting of the antisymmetric superposition of two of
the\ fundamental gap solitons previously studied.

%%%%%%%%%%%%%%%%%%%%%%%%%%%%%%%%%%%%%%%%%%%%%%%%%%%%%%%%%%%%%%%%%%%%%%%%%%%%%%

%%%%%%%%%%%%%%%%%%%%%%%%%%%%%%%%%%%%%%%%%%%%%%%%%%%%%%%%%%%%%%%%%%%%%%%%%%%%%%

The lower trajectories in Fig. \ref{Fig5}, which bifurcate from the second 1D
Bloch band, constitute a family that has been referred to as the
\emph{subfundamental} family \cite{Mal1}. Gap solitons in this family are
associated with the first excited ($n=2$) energy band of the corresponding 1D
axial problem, which reflects in the fact that they exhibit two major peaks in
the axial direction (\textit{i.e.}, one axial node) localized in a single
lattice site (see the lower inset in Fig. \ref{Fig5} as well as panels F in
Fig. \ref{Fig6}). Unlike the previous case, these solitons (which can be
regarded as excited states featuring an embedded dark soliton) are not bound
states of two fundamental $(1,0,0)$ gap solitons. They are fundamental
solitons of the $(2,0,0)$ family which can be used as elementary constituents
of more complex compound structures.

%%%%%%%%%%%%%%%%%%%%%%%%%%%%%%%%%%%%%%%%%%%%%%%%%%%%%%%%%%%%%%%%%%%%%%%%%%%%%%

%%%%%%%%%%%%%%%%%%%%%%%%%%%%%%%%%%%%%%%%%%%%%%%%%%%%%%%%%%%%%%%%%%%%%%%%%%%%%%

As is apparent from the above results, the predictions from the 1D effective
model (\ref{E1DE}) are always in good quantitative agreement with the 3D
numerical results.

\section{IV. STABILITY ANALYSIS}

Stability is of primary importance for the experimental relevance of gap
solitons. In this Section we are interested in determining whether the
effective equation (\ref{E1DE}) is able to correctly predict the stability
properties of the gap solitons previously considered. To this end we perform
a stability analysis based on the above effective equation and compare with 
the respective 3D results. We begin by deriving the Bogoliubov--de Gennes 
(BdG) equations corresponding to Eq. (\ref{E1DE}) and then obtain 
numerically the frequency spectrum of the elementary excitations.
To derive the BdG equations, we perturb a given gap soliton stationary
wavefunction $\phi_{0}(z)$ in the form \cite{Carre1}
\begin{equation}
\phi(z,t)=\left[  \phi_{0}(z)+u(z)e^{-i\omega t}+v^{*}(z)e^{i\omega t}\right]
e^{-i\mu t/\hbar} \label{BDG1}%
\end{equation}
and substitute in Eq. (\ref{E1DE}) retaining only linear terms in the
amplitudes $u$ and $v$ of the normal modes of the system. The numerical
solution of the resulting linear equations provides the frequencies of the
elementary excitations determining the dynamical stability of the soliton. A 
similar procedure has been followed to derive the BdG equations corresponding
to the 3D GPE (\ref{3DGPE}).

\begin{figure}[t]
\begin{center}
\includegraphics[width=8.5cm]{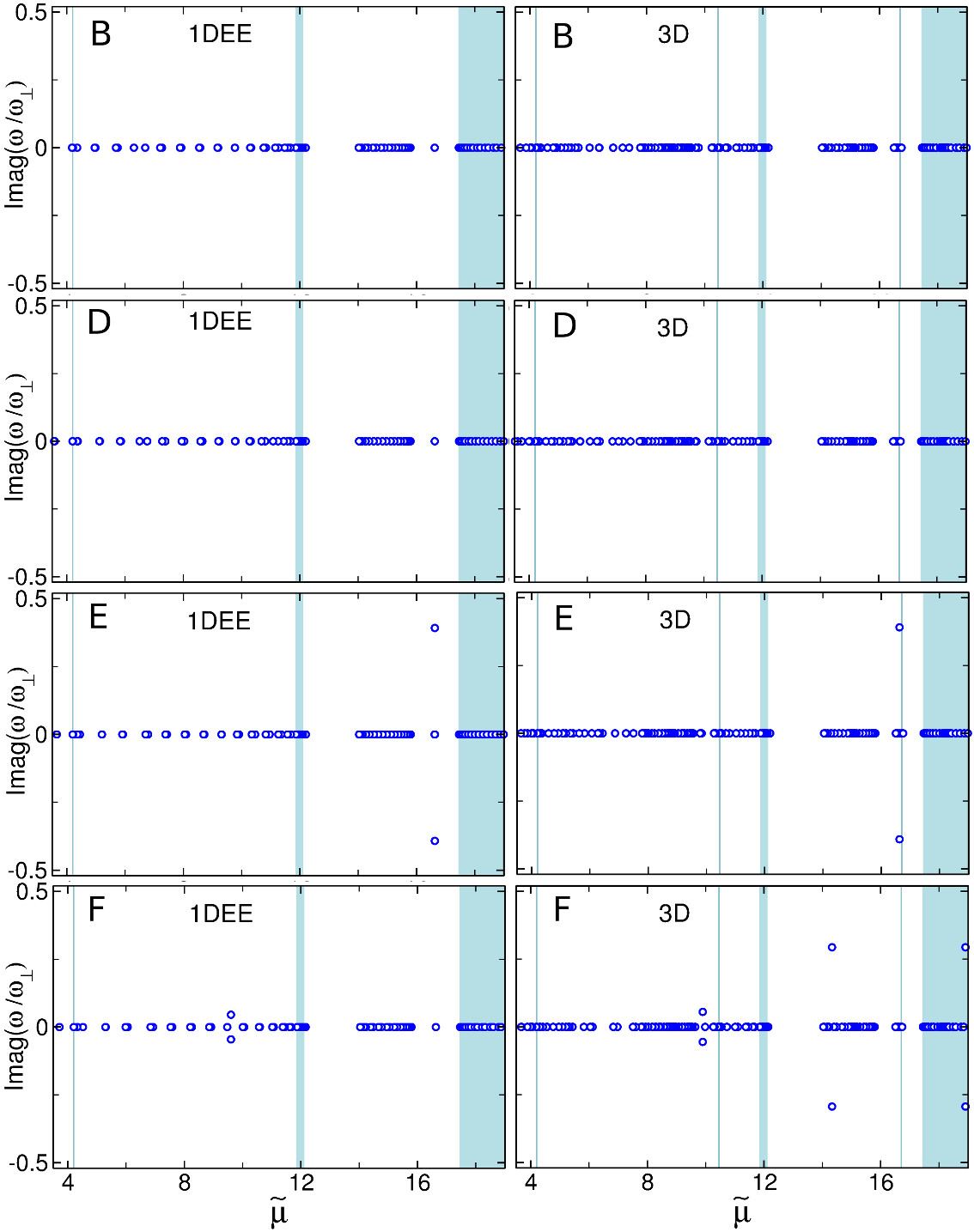}
\end{center}
\caption{(Color online) Panels B, D, E, and F show the frequency spectra of
the elementary excitations of the gap solitons labeled with the same letters
as follow from the BdG equations corresponding to the 1D effective model
(\ref{E1DE}) (left panels) and to the 3D GPE (right panels). The horizontal
axis in each figure represents the real part of the excitation frequencies
$\omega$ shifted by the chemical potential $\tilde{\mu}_{0}$ of the
corresponding stationary soliton, $\tilde{\mu}=\tilde{\mu}_{0}%
+\mathrm{\operatorname{Re}}(\omega/E_{R})$, while the vertical axes show the
respective imaginary parts, $\mathrm{\operatorname{Im}}(\omega/\omega_{\bot}%
)$. Note that in all cases considered $\tilde{\mu}_{0}=16.6$.}%
\label{Fig7}%
\end{figure}

Results for gap solitons B, D, E, and F are given in the panels labeled with
the same letters in Fig. \ref{Fig7}.
%
%%%%%%%%%%%%%%%%%%%%%%%%%%%%%%%%%%%%%%%%%%%%%%%%%%%%%%%%%%%%%%%%%%%%%%%%%%%%%%
%%%%%%%%%%%%%%%%%%%%%%%%%%%%%%%%%%%%%%%%%%%%%%%%%%%%%%%%%%%%%%%%%%%%%%%%%%%%%%
%%%%%%%%%%%%%%%%%%%%%%%%%%%%%%%%%%%%%%%%%%%%%%%%%%%%%%%%%%%%%%%%%%%%%%%%%%%%%%
\begin{figure}[t]
\begin{center}
\includegraphics[width=8.5cm]{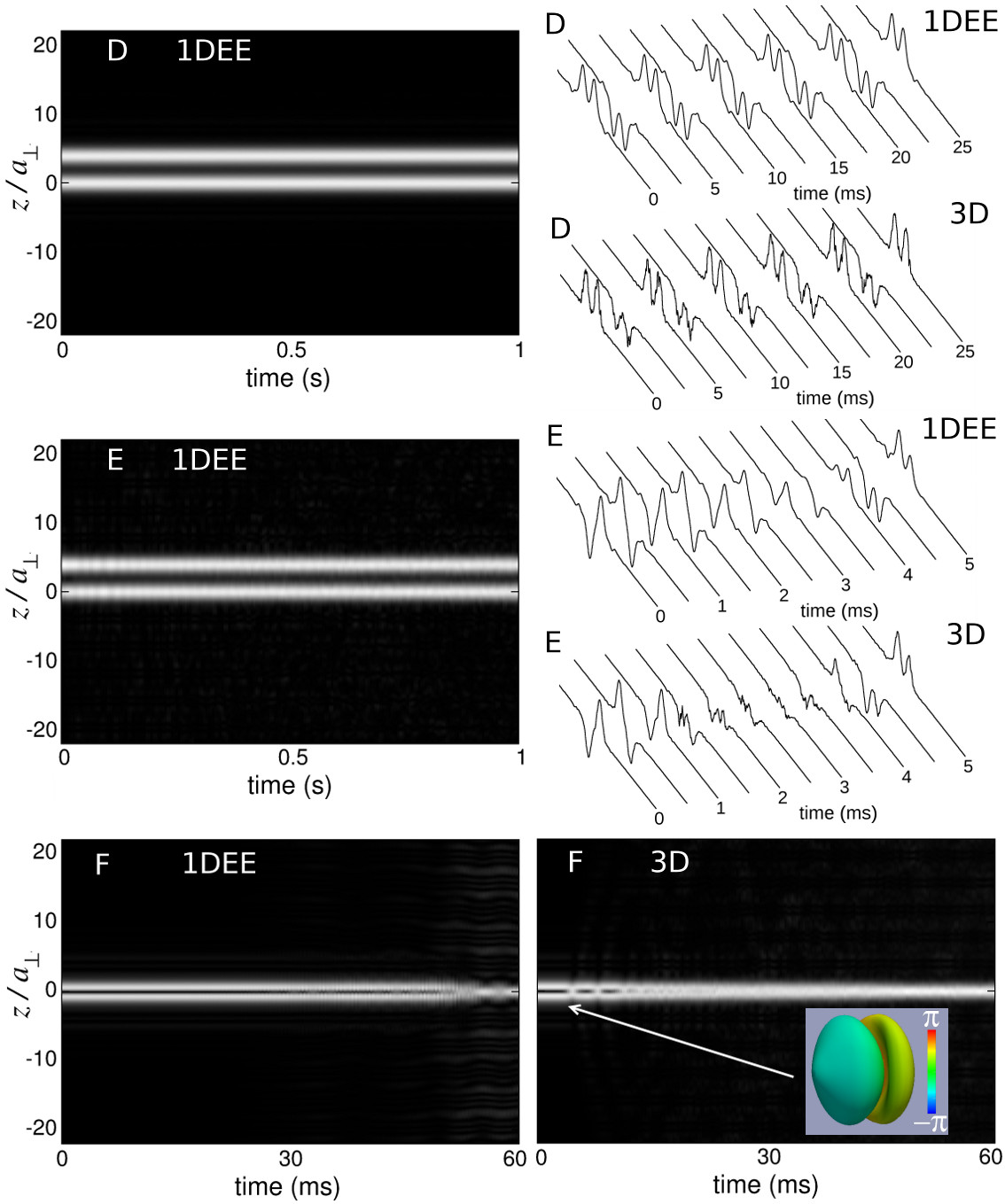}
\end{center}
\caption{(Color online) Representative examples showing the long time 
evolution of the stationary solitons D, E, and F after a random small 
perturbation (see text for details).}%
\label{Fig8}%
\end{figure}
%%%%%%%%%%%%%%%%%%%%%%%%%%%%%%%%%%%%%%%%%%%%%%%%%%%%%%%%%%%%%%%%%%%%%%%%%%%%%%
Left panels show the corresponding excitation spectra as obtained from the 1D
effective model (\ref{E1DE}), while right panels display the eigenfrequencies
of the BdG equations associated to the 3D GPE. The horizontal axis in each
figure represents the real part of the excitation frequencies $\omega$ shifted
by the chemical potential $\tilde{\mu}_{0}$ of the corresponding stationary
soliton, $\tilde{\mu}=\tilde{\mu}_{0}+\mathrm{\operatorname{Re}}(\omega
/E_{R})$. Vertical axes show the respective imaginary parts,
$\mathrm{\operatorname{Im}}(\omega/\omega_{\bot})$. As is well known, the
absence of complex eigenfrequencies in the spectrum implies the linear
stability of the system. Of course, the excitation spectrum of the full 3D
problem necessarily contains frequencies that cannot exist in the effective
problem with reduced dimensionality (for instance, those associated to radial
normal modes). This is an expected result. The point, however, is that, except
in those cases where the solitons exhibit a nontrivial (unstable) transverse
configuration, one might also reasonably expect that the stability properties
of the different gap solitons will be dictated by the axial degrees of
freedom, which ultimately are the ones responsible for the existence of
self-trapping. If this expectation proves true, the effective 1D model
(\ref{E1DE}) could be a reliable tool for the investigation of the stability
of these nonlinear structures. As is apparent from a comparison between left
and right panels in Fig. \ref{Fig7}, this is indeed the case: except for the
gap soliton F (lower panels), which is a bright soliton containing a dark
soliton that extends considerably in the radial direction, the effective 1D
model predicts accurately the complex eigenfrequencies determining the
stability of the various gap solitons. According to the excitation spectra
shown in the figure, the fundamental soliton B and the symmetric composite
soliton D are linearly stable while soliton E (the antisymmetric counterpart
of the latter) is unstable. 
Since in all cases considered $\tilde{\mu}_{0}=16.6$, it is clear from the 
figure that the spectrum of this soliton exhibits a pair of purely 
imaginary eigenfrequencies, indicating an exponential instability.
As for soliton F, even though the effective model (\ref{E1DE}) still predicts 
correctly the (oscillatory) unstable nature of this soliton, it cannot give a 
quantitative account of its 3D excitation spectrum which now contains a 
quartet of complex frequencies not appearing in the corresponding 
1D spectrum (which exhibits another quartet of complex frequencies, although
only a pair is visible in the field of view of the figure).
One expects these complex frequencies not appearing in the 1D spectrum to be 
associated with unstable transverse excitations of the embedded dark soliton 
(snake-like instability). To verify this and the previous conclusions we have 
introduced an additive Gaussian white noise to produce a random small 
perturbation in the stationary configuration of the different gap solitons and 
have followed the ensuing time evolution by integrating numerically both the 
effective 1D equation (\ref{E1DE}) and the 3D GPE (\ref{3DGPE}). The results 
obtained are collected in Fig. \ref{Fig8}.
The upper left panel shows the dynamical evolution of soliton D in terms of a
density map where brighter regions correspond to higher densities. This image,
which has been obtained from the 1D effective equation (\ref{E1DE}), is
indistinguishable from the corresponding 3D result and confirms the stability
of the symmetric composite soliton D as predicted by the linear stability
analysis of Fig. \ref{Fig7}. The same holds true for the fundamental soliton B
(not shown in the figure). For later convenience, the upper right panel
displays the time evolution of the real part of the wavefunction of soliton D
(in arbitrary units) obtained from both the 1D (top) and the 3D models
(bottom). Since this soliton is stable, its dynamical evolution, apart from
small local fluctuations induced by the Gaussian perturbation, is given
essentially by a periodic amplitude modulation $\mathrm{\operatorname{Re}%
}(\phi(t))\simeq\phi(0)\cos(\mu t/\hbar)$ with period $T\simeq0.2$ ms, as
corresponds to a (quasi) stationary state. Thus stability implies, in
particular, that the relative phase between any pair of points in the global
amplitude envelope must remain (quasi) constant over time. Accordingly, even
though an analysis of the density evolution of the antisymmetric composite
soliton E (middle left panel in Fig. \ref{Fig8}) seems to indicate that this
soliton is stable (the same conclusions follow from both 1D and 3D results), a
more detailed study including also phase information reveals that this is not
true. Indeed, as can be seen from the middle right panel, in this case the
time evolution of the real part of the wavefunction no longer is a periodic
amplitude modulation. While the two major peaks are initially out of phase,
both the 1D effective model (top) and the 3D model (bottom) predict that they
become essentially in phase at $t\approx4.5-5$ ms, so that, this soliton turns
out to be unstable, in agreement with the linear stability analysis of Fig.
\ref{Fig7}. This example clearly reflects the importance of the spectrum of
elementary excitations in the analysis of the dynamical stability of these
nonlinear structures.

In the two lower panels we compare the evolution in time of the density of
soliton F obtained from Eq. (\ref{E1DE}) (left panel) with that obtained from
the 3D GPE (right panel). As is apparent, in this case the effective 1D model
predicts a longer soliton lifetime than the 3D model, in good agreement\ with
the linear stability analysis of Fig. \ref{Fig7}. The inset in the right panel
shows that the soliton decay occurs when the nodal plane of the embedded dark
soliton begins to undergo a considerable deformation, thus, indicating that
the decay is a consequence of a transverse modulational instability which
clearly cannot be accounted for with a 1D model.

One may wonder why the effective model (\ref{E1DE}) is able to properly 
predict the stability properties of the gap solitons even though, in the 
presence of the optical lattice (and because of the high frequency axial 
motion it might induce), in general, it is unable to accurately reproduce 
the detailed evolution in time. To address this question, we begin by noting 
that the adiabatic (Born--Oppenheimer) approximation essentially consists in 
assuming that the axial dynamics is so slow in comparison to the typical time 
scale of the radial degrees of freedom that, at every instant $t$, the latter 
can adjust to their equilibrium configuration compatible with the axial 
configuration $n_{1}$ occurring at that instant. This is exactly what the 
stationary radial equation (\ref{II-7}) reflects. In the presence of an 
optical lattice, one thus reasonably expects this approximation to work better 
in those situations where the time variable does not play a prominent role, 
such as occurs for the stationary solutions of the time-dependent GPE, which 
are completely determined from a (time-independent) eigenvalue problem.
Something similar can be expected for the linear stability properties of the 
system. Indeed, such properties are completely characterized by the frequency 
spectrum of the elementary excitations, which follow from the solution of an 
eigenvalue problem (BdG equations) where time does not play any role. 
Even though an instability manifests itself in the evolution in time, the very 
existence of a certain unstable mode is a characteristic feature of the system 
which depends only on the interplay among the various energies and interactions 
contributing to the Hamiltonian. This explains why the effective model 
(\ref{E1DE}) can predict the stability properties of the gap solitons even 
though it is unable to properly account for their dynamical evolution.

\section{V. CONCLUSION}

Most theoretical studies of gap solitons realized in condensates loaded in 1D
optical lattices have been carried out in terms of the 1D GPE. This is
motivated, in part, by the analogy of the latter equation with the 1D NLSE
governing the evolution of similar structures in nonlinear optics and, in
part, by the fact that, as occurs with dark solitons, one would reasonably
expect these structures to be more stable in a quasi-1D regime, where their
potential decay would be inhibited by the strong radial confinement. However,
it has been demonstrated that robust, long-lived gap solitons exist in a 3D
regime where many higher-order radial modes are excited \cite{GS-3D}.
Moreover, matter-wave gap solitons are intrinsically 3D and in realistic
situations can hardly satisfy conditions (\ref{con3}), which are necessary for
the validity of the 1D GPE. In these circumstances, effective equations like
Eq. (\ref{E1DE}), which can take the 3D character of stationary
BECs into account by incorporating contributions from higher-order radial
modes, offer a clear advantage over the latter. However, the question remains
open whether the effective 1D equation (\ref{E1DE}) represents a reliable
alternative to a fully 3D treatment. In this work we have addressed this
question by analyzing the ability of the above effective equation to reproduce
the fundamental physical properties of stationary matter-wave gap solitons in
realistic conditions. Our results demonstrate that the predictions from the
effective model (\ref{E1DE}) are in good quantitative agreement with the exact
numerical results from the full 3D GPE. In particular, unlike the standard 1D
GPE (which fails in most cases of practical interest), the 1D effective model
(\ref{E1DE}) correctly predicts the distinctive trajectories characterizing
the different gap soliton families as well as the corresponding axial
wavefunctions along the entire band gaps. Our results also show that this
effective model can predict correctly the stability properties of the
different gap soliton families. This follows from both a linear stability
analysis in terms of the corresponding BdG equations and a representative 
set of numerical computations monitoring the long time response of the 
system to a sudden small random perturbation. 
In particular, by numerically solving the BdG equations we have
analyzed the prediction the effective model makes for the spectrum of
elementary excitations, which proves to be essential for unambiguously
determining the stability properties of certain matter-wave gap solitons. Our
results show that, except in those cases where the soliton features a
nontrivial (unstable) transverse configuration, the model is able to correctly
predict the spectrum of complex eigenfrequencies responsible for the dynamical
stability of the system. These results thus indicate that the above effective
1D model can be a useful tool for the physical description of realistic
matter-wave gap solitons in 1D optical lattices.

As a natural extension of the present work, one might consider to quantitatively
investigate the effects of quantum fluctuations on the above stationary 
matter-wave gap solitons. In the tight-binding regime ($s\gg1$), the tunneling 
of atoms between adjacent lattice sites can be strongly inhibited. Under these 
conditions, intersite atom-number fluctuations become significantly suppressed 
while the corresponding phase fluctuations become enhanced, which can lead to a 
reduced relative phase-coherence. While for fundamental gap solitons (those 
located at a single lattice well) with a few hundreds atoms these quantum effects 
are not expected to be very significant, they can be more important in the case 
of composite in-phase or out-of-phase solitons. The problem of quantifying the 
contribution of these quantum fluctuations can be addressed in terms of a 
stochastic phase-space description based on the Truncated Wigner Approximation 
(TWA) \cite{TWA}. In this respect, we note that since Eq. (\ref{E1DE}) properly 
captures the Bogoliubov excitation spectrum of the system, it represents a 
particularly convenient starting point for such an analysis. This will be the 
subject of a future publication.


\begin{thebibliography}{99}                                                                                               %


\bibitem {PeSmB}C. J. Pethick and H. Smith, \textit{Bose--Einstein Condensation
in Dilute Gases}, 2nd ed. (Cambridge University Press, Cambridge, 2008).

\bibitem {PiStB}L. Pitaevskii and S. Stringari, \textit{Bose--Einstein
Condensation} (Clarendon Press, Oxford, 2003).

\bibitem {Carre1}R. Carretero-Gonz\'{a}lez, D. J. Frantzeskakis, and P. G.
Kevrekidis, Nonlinearity \textbf{21}, R139 (2008).

\bibitem {RevDS}D J Frantzeskakis, J. Phys. A: Math. Theor. \textbf{43},
213001 (2010).

\bibitem {Kon1}V. A. Brazhnyi and V. V. Konotop, Mod. Phys. Lett. B
\textbf{18}, 627 (2004).

\bibitem {Mor1}O. Morsch and M. Oberthaler, Rev. Mod. Phys. \textbf{78}, 179 (2006).

\bibitem {KivGOP}Y. S. Kivshar and G. P. Agrawal, \textit{Optical Solitons:
From Fibers to Photonic Crystals} (Academic, San Diego, 2003).

\bibitem {Kiv2D}E. A. Ostrovskaya and Y. S. Kivshar, Phys. Rev. Lett.
\textbf{90}, 160407 (2003).

\bibitem {RMP-Malomed}Y. V. Kartashov, B. A. Malomed, and L. Torner, Rev. Mod.
Phys. \textbf{83}, 247 (2011).

\bibitem {Burger1}S. Burger, K. Bongs, S. Dettmer, W. Ertmer, K. Sengstock, A.
Sanpera, G. V. Shlyapnikov, and M. Lewenstein, Phys. Rev. Lett. \textbf{83},
5198 (1999).

\bibitem {Phil1}J. Denschlag, J. E. Simsarian, D. L. Feder, Charles W. Clark,
L. A. Collins, J. Cubizolles, L. Deng, E. W. Hagley, K. Helmerson, W. P.
Reinhardt, S. L. Rolston, B. I. Schneider, and W. D. Phillips, Science
\textbf{287}, 97 (2000).

\bibitem {Hau1}Z. Dutton, M. Budde, C. Slowe, and L. V. Hau, Science
\textbf{293}, 663 (2001).

\bibitem {Corn1}B. P. Anderson, P. C. Haljan, C. A. Regal, D. L. Feder, L. A.
Collins, C. W. Clark, and E. A. Cornell, Phys. Rev. Lett. \textbf{86}, 2926 (2001).

\bibitem {DS-2}P. Engels and C. Atherton, Phys. Rev. Lett. \textbf{99}, 160405 
(2007); C. Becker, S. Stellmer, P. Soltan-Panahi, S. D\"{o}rscher, M. Baumert, 
E.-M. Richter, J. Kronj\"{a}ger, K. Bongs, and K. Sengstock, Nature Physics 
\textbf{4}, 496 (2008); S. Stellmer, C. Becker, P. Soltan-Panahi, E.-M. Richter, 
S. D\"{o}rscher, M. Baumert, J. Kronj\"{a}ger, K. Bongs, and K. Sengstock,
Phys. Rev. Lett. \textbf{101}, 120406 (2008); A. Weller, J. P. Ronzheimer, C. Gross, 
J. Esteve, M. K. Oberthaler, D. J. Frantzeskakis, G. Theocharis, and P. G. 
Kevrekidis, \textit{ibid}. \textbf{101}, 130401 (2008); I. Shomroni, E. Lahoud, 
S. Levy, and J. Steinhauer, Nature Physics \textbf{5}, 193 (2009).

\bibitem {Hul1}K. E. Strecker, G. B. Partridge, A. G. Truscott, and R. G.
Hulet, Nature \textbf{417}, 150 (2002).

\bibitem {Salo1}L. Khaykovich, F. Schreck, G. Ferrari, T. Bourdel, J.
Cubizolles, L. D. Carr, Y. Castin, C. Salomon, Science \textbf{296}, 1290 (2002).

\bibitem {Wie1}S. L. Cornish, S. T. Thompson, and C. E. Wieman, Phys. Rev.
Lett. \textbf{96}, 170401 (2006).

\bibitem {Ober1}B. Eiermann, Th. Anker, M. Albiez, M. Taglieber, P. Treutlein,
K.-P.Marzlin, and M. K. Oberthaler, Phys. Rev. Lett. \textbf{92}, 230401 (2004).

\bibitem {GS1D}F. Kh. Abdullaev, B. Baizakov, S. Darmanyan, V. Konotop and M.
Salerno, Phys. Rev. A \textbf{64}, 043606 (2001).

\bibitem {Kiv1}P. J. Y. Louis, E. A. Ostrovskaya, C. M. Savage, and Y. S.
Kivshar, Phys. Rev. A \textbf{67}, 013602 (2003).

\bibitem {Efre1}N. K. Efremidis and D. N. Christodoulides, Phys. Rev. A
\textbf{67}, 063608 (2003).

\bibitem {Abd1}F. Kh. Abdullaev and M. Salerno, Phys. Rev. A \textbf{72},
033617 (2005).

\bibitem {Mal1}T. Mayteevarunyoo and B. A. Malomed, Phys. Rev. A \textbf{74},
033616 (2006).

\bibitem {Wu1}Y. Zhang and B. Wu, Phys. Rev. Lett. \textbf{102}, 093905
(2009); Y. Zhang, Z. Liang, and B. Wu, Phys. Rev. A \textbf{80}, 063815 (2009).

%================================================================================

\bibitem {Das1}K. K. Das, Phys. Rev. A \textbf{66}, 053612 (2002).

\bibitem {VDB1}A. Mu\~{n}oz Mateo and V. Delgado, Phys. Rev. A \textbf{74},
065602 (2006).

\bibitem {VDB2}A. Mu\~{n}oz Mateo and V. Delgado, Phys. Rev. A \textbf{75},
063610 (2007).

\bibitem {Fuch1}J. N. Fuchs, X. Leyronas, and R. Combescot, Phys. Rev. A 
\textbf{68}, 043610 (2003); F. Gerbier, Europhys. Lett. \textbf{66}, 771 (2004).

\bibitem {Jack1}A. D. Jackson, G. M. Kavoulakis, and C. J. Pethick, Phys. Rev.
A \textbf{58}, 2417 (1998).

\bibitem {Yuka1}K.-P. Marzlin and V. I. Yukalov, Eur. Phys. J. D \textbf{33},
253 (2005).

\bibitem {Nico3}A. Bala\v{z}, R. Paun, A. I. Nicolin, S. Balasubramanian, and 
R. Ramaswamy, Phys. Rev. A \textbf{89}, 023609 (2014).

\bibitem {Salas1}L. Salasnich, A. Parola, and L. Reatto, Phys. Rev. A
\textbf{65}, 043614 (2002); L. Salasnich, Laser Phys. \textbf{12}, 198 (2002).

\bibitem {Ef1DEqs}A. Mu\~{n}oz Mateo and V. Delgado, Phys. Rev. A \textbf{77}, 
013617 (2008); Ann. Phys. \textbf{324}, 709 (2009).

\bibitem {Adhik1}C. A. G. Buitrago and S. K. Adhikari, J. Phys. B: At. Mol.
Opt. Phys. \textbf{42}, 215306 (2009).

\bibitem {Kam1}A. M. Kamchatnov and V. S. Shchesnovich, Phys. Rev. A
\textbf{70}, 023604 (2004).

\bibitem {Clark1}M. Edwards, L. M. DeBeer, M. Demenikov, J. Galbreath, T. J.
Mahaney, B. Nelsen, and C. W. Clark, J. Phys. B: At. Mol. Opt. Phys.
\textbf{38}, 363 (2005).

\bibitem {Modug1}P. Massignan and M. Modugno, Phys. Rev. A \textbf{67}, 023614
(2003).

\bibitem {You1}W. Zhang and L. You, Phys. Rev. A \textbf{71}, 025603 (2005).

\bibitem {Salas3}L. Salasnich, B. A. Malomed, and F. Toigo, Phys. Rev. A
\textbf{76}, 063614 (2007).

\bibitem {Salas4}L. Salasnich, J. Phys. A: Math. Theor. \textbf{42}, 335205 (2009).

\bibitem {Adhik2}L. E. Young S., L. Salasnich, and S. K. Adhikari, Phys. Rev.
A \textbf{82}, 053601 (2010).

\bibitem {Carre3}S. Middelkamp, J. J. Chang, C. Hamner, R.
Carretero-Gonz\'{a}lez, P. G. Kevrekidis, V. Achilleos, D. J. Frantzeskakis,
P. Schmelcher, and P. Engels, Phys. Lett. A \textbf{375}, 642 (2011).

\bibitem {Adhik4}P. Muruganandam, and S. K. Adhikari, Laser Phys. \textbf{22},
813 (2012).

\bibitem {Salas6}L. Salasnich and B. A. Malomed, Phys. Rev. A \textbf{87},
063625 (2013).

\bibitem {Salas5}L. Salasnich and B. A. Malomed, J. Phys. B: At. Mol. Opt.
Phys. \textbf{45}, 055302 (2012).

\bibitem {Mal2}W. B. Cardoso, J. Zeng, A. T. Avelar, D. Bazeia, and B. A.
Malomed, Phys. Rev. E \textbf{88}, 025201 (2013).

\bibitem {Ben1}T. Yang, A. J. Henning, and K. A. Benedict, J. Phys. B: At.
Mol. Opt. Phys. \textbf{47}, 035302 (2014).

%==================================================================================

\bibitem {Kev1}S. Middelkamp, G. Theocharis, P. G. Kevrekidis, D. J.
Frantzeskakis, and P. Schmelcher, Phys. Rev. A \textbf{81}, 053618 (2010).

\bibitem {Wel1}G. Theocharis, A. Weller, J. P. Ronzheimer, C. Gross, M. K.
Oberthaler, P. G. Kevrekidis, and D. J. Frantzeskakis, Phys. Rev. A
\textbf{81}, 063604 (2010).

\bibitem {Kev2}C. Wang, P. G. Kevrekidis, T. P. Horikis, D. J. Frantzeskakis,
Phys. Lett. A \textbf{374}, 3863 (2010).

\bibitem {Isa1}J. Armijo, T. Jacqmin, K. Kheruntsyan, and I. Bouchoule, Phys.
Rev. A \textbf{83}, 021605(R) (2011); J. Armijo, Phys. Rev. Lett.
\textbf{108}, 225306 (2012).

\bibitem {VDB7}A. Mu\~{n}oz Mateo, V. Delgado, and B. A. Malomed, Phys. Rev. A
\textbf{83}, 053610 (2011).

\bibitem {PRL06}A. Mu\~{n}oz Mateo and V. Delgado, Phys. Rev. Lett.
\textbf{97}, 180409 (2006).

\bibitem {Mott}J. A. Huhtam\"{a}ki, M. M\"{o}tt\"{o}nen, T. Isoshima, V.
Pietil\"{a}, and S. M. Virtanen, Phys. Rev. Lett. \textbf{97}, 110406 (2006).

\bibitem {Arim1}M. Cristiani, O. Morsch, J. H. M{\"u}ller, D. Ciampini, and E.
Arimondo, Phys. Rev. A \textbf{65}, 063612 (2002).

\bibitem {Ingus1}N. Fabbri, D. Cl\'{e}ment, L. Fallani, C. Fort, M. Modugno,
K. M. R. van der Stam, and M. Inguscio, Phys. Rev. A \textbf{79}, 043623 (2009).

\bibitem {GS-3D}A. Mu\~{n}oz Mateo, V. Delgado, and B. A. Malomed, Phys. Rev.
A \textbf{82}, 053606 (2010).

\bibitem {VDB8}A. Mu\~{n}oz Mateo, V. Delgado, Phys. Rev. E \textbf{88},
042916 (2013).

\bibitem {TWA}J. Ruostekoski and A. D. Martin, Truncated Wigner method 
for Bose gases. In, Quantum Gases: Finite Temperature and Non-Equilibrium 
Dynamics (Vol. 1 Cold Atoms Series), Pag. 203. N. Proukakis, S. Gardiner, 
M. Davies, M. Szymanska, and N. Nygaard (eds.) (Imperial College Press, 
London, 2013).


\end{thebibliography}
\end{document}